\documentclass[aps,superscriptaddress,twocolumn, prl]{revtex4-1}
\usepackage{amsmath}
\usepackage{amsfonts}
\usepackage{amssymb, bbm}
\usepackage{amsthm}
\usepackage{fullpage}
\usepackage{graphicx}

\usepackage{tikz}
\usetikzlibrary{patterns}
\usetikzlibrary{calc,decorations.pathreplacing}
\usetikzlibrary{arrows,shapes}
\usepackage{subfigure}
\usepackage{changes}

\newcommand{\be}{\begin{equation}}
\newcommand{\ee}{\end{equation}}
\newcommand{\cH}{\mathcal{H}}

\newcommand{\cS}{\mathcal{S}}
\newcommand{\cE}{\mathcal{E}}
\newcommand{\aQ}{\tilde{\mathcal{Q}}}
\newcommand{\id}{\mathbbm{I}}

\newcommand{\C}{\mathbbm{C}}

\newcommand{\ket}[1]{\left| #1 \right\rangle}
\newcommand{\bra}[1]{\left\langle #1 \right|}
\newcommand{\proj}[1]{\ket{#1}\!\bra{#1}}

\def\braket#1#2{\langle#1|#2\rangle}

\let\OLDthebibliography\thebibliography
\renewcommand\thebibliography[1]{
  \OLDthebibliography{#1}
  \setlength{\parskip}{0pt}
  \setlength{\itemsep}{0pt plus 0.3ex}
}

\begin{document}
\title{Almost quantum correlations violate the no-restriction hypothesis}
\author{Ana Bel\'en Sainz}\affiliation{Perimeter Institute for Theoretical Physics, 31 Caroline St. N, Waterloo, Ontario, Canada, N2L 2Y5}
\author{Yelena Guryanova}\affiliation{Institute for Quantum Optics and Quantum Information (IQOQI), Boltzmanngasse 3 1090, Vienna, Austria}
\author{Antonio Ac\'in}\affiliation{ICFO-Institut de Ciencies Fotoniques, The Barcelona Institute of Science and Technology, 08860 Castelldefels (Barcelona), Spain}\affiliation{ICREA-Instituci\'o Catalana de Recerca i Estudis Avan\c{c}ats, 08010 Barcelona, Spain}
\author{Miguel Navascu\'es}\affiliation{Institute for Quantum Optics and Quantum Information (IQOQI), Boltzmanngasse 3 1090, Vienna, Austria}

\date{\today}

\begin{abstract}
To identify which principles characterize quantum correlations, it is essential to understand in which sense this set of correlations differs from that of almost quantum correlations. We solve this problem by invoking the so-called no-restriction hypothesis, an explicit and natural axiom in many reconstructions of quantum theory stating that the set of possible measurements is the dual of the set of states. We prove that, contrary to quantum correlations, no generalised probabilistic theory satisfying the no-restriction hypothesis is able to reproduce the set of almost quantum correlations. Therefore, any theory whose correlations are exactly, or very close to, the almost quantum correlations necessarily requires a rule limiting the possible measurements. Our results suggest that the no-restriction hypothesis may play a fundamental role in singling out the set of quantum correlations among other non-signalling ones.
\end{abstract}
\maketitle

One of the most pronounced phenomena to reveal that Nature is not classical is Bell nonlocality.
Indeed, quantum realisations of the so-called `Bell experiment' have confirmed that Bell inequality violations can be observed. The study of quantum nonlocality has hence become an active field of research aimed, on the one hand, at exploiting these correlations with no classical analogue and, on the other, at understanding the counter-intuitive features of quantum theory, which represents our most accurate description of nature to date.

Part of this research has focused on deriving the correlations produced by quantum mechanics solely from physical principles, i.e. without making reference to the underlying mathematical structure of Hilbert spaces, vectors, self-adjoint operators and so forth. This program was initiated by Popescu and Rohrlich \cite{popescu} (see also \cite{tsirel_box}), who showed that relativistic considerations (no-signalling faster than light) alone are not enough to single out the quantum set, and in fact allow two spatially separated parties to generate correlations impossible to approximate in quantum theory. 

More than a decade later, Van Dam showed that the existence of some supra-quantum correlations compatible with special relativity could have implausible consequences from an information processing point of view \cite{jeanclaude}. Since then, remarkable progress has been made in identifying physical principles that constrain non-signalling correlations \cite{brassard, linden, marcin, mac_loc, loc_orth}. However, all these attempts turned out to be insufficient in singling out quantum correlations, since (with the possible exception of \cite{marcin}), they proved to be satisfied by a set of correlations slightly larger than the quantum one, called \textit{almost quantum} \cite{almost}. This opens the question of whether there is a fundamental limitation to this research program, or if it is the case that almost quantum (AQ) correlations could really be the correlations allowed by a theory alternative to quantum mechanics. 

In this work we discuss the properties that a physical theory should have in order to predict AQ correlations. For this we work within the framework of generalized probabilistic theories (GPTs). We find that should such a theory exist, it will not satisfy the no-restriction hypothesis, in the sense that {the measurements allowed on the systems should be further constrained}. 
Indeed, the no-restriction hypothesis imposes that any mathematically well defined measurement in the theory should be physically allowed. Depending on the interpretation, our results then provide a means to argue that AQ correlations may not be physical after all, or on the contrary, highlight how restrictive the no-restriction hypothesis is. 

\textit{Bell scenarios and the almost quantum set.--} In a Bell scenario, a set of distant parties perform space-like separated actions on their share of a system. Each party can choose from $m$ different measurements to perform on their subsystem, obtaining one of $d$ possible outcomes. Such a Bell scenario is referred to as $(n,m,d)$, where $n$ denotes the total number of parties. 
For each party $k \in \{1, \ldots, n\}$, we use $x_k$ to denote their choice of measurement setting and $a_k$ to denote the outcome obtained. After performing measurements on many independent copies of the system, the parties can estimate the conditional probability distribution (also called behavior) $p(a_1 \ldots a_n | x_1 \ldots x_n)$. These correlations are the objects we aim to characterise.

The correlations $p(a_1 \ldots a_n | x_1 \ldots x_n)$ that the parties can observe are limited by the physical theory modeling the behavior of their experimental apparatuses. 
Quantum theories allow correlations of the form
\be
p(a_1 \ldots a_n | x_1 \ldots x_n) = \bra{\Psi}\otimes_{k=1}^n M^{(k)}_{a_k|x_k}\ket{\Psi},
\label{born}
\ee
\noindent where $\Psi\in \cH_1\otimes...\otimes \cH_n$ is a normalized vector; $\cH_1,...,\cH_n$ are arbitrary Hilbert spaces and, for all $k,x$,  $\{M^{(k)}_{a|x}\}_a$ is a positive operator valued measure (POVM) acting on $\cH_k$.

The set of AQ correlations, denoted by $\aQ$, can be mathematically defined in many equivalent ways. Within the language of quantum theory, a behaviour $p(a_1 \ldots a_n | x_1 \ldots x_n)$ has an AQ realisation if there exists a Hilbert space $\cH$, projective measurements $\{N^{(k)}_{a_k|x_k}\}_{a_k}$ acting on $\cH$ and a quantum state $\ket{\Psi}$ such that 
\begin{enumerate}
\item
$p(a_1,...,a_n|x_1,...x_n)=\bra{\Psi}N^{(1)}_{a_1|x_1}...N^{(n)}_{a_n|x_n}\ket{\Psi}$.
\item
$\prod_{k=1}^n N^{(k)}_{a_k|x_k} \ket{\Psi} = \prod_{k=1}^n N^{(\pi(k))}_{a_{\pi(k)}|x_{\pi(k)}} \ket{\Psi}$, for all permutations $\pi$ of the parties. 
\end{enumerate}
As shown in \cite{almost}, all correlations of the form (\ref{born}) are AQ. However, even in the simplest Bell scenarios, there exist instances of AQ correlations which do not admit a quantum realization.

It then follows that all quantum correlations belong to the AQ set. As shown in \cite{almost}, this inclusion is strict. Moreover, contrary to the quantum case, the problems of testing membership in the AQ set as well as computing maximal violations of Bell inequalities therein can be solved efficiently via semidefinite programming (SDP) \cite{sdp}.

\textit{Generalised Probabilistic Theories.--} In order to explore the properties of a hypothetical theory that describes AQ correlations, we make use of the formalism of generalised probabilistic theories (GPTs), also referred to as the convex operational framework. 

A GPT is specified by a list of system types, together with composition rules specifying which system type describes the combination of several other types. In a GPT, each system is described by a \textit{state} $\Psi$ which is fully specified by the vector of probabilities for the outcomes of all measurements that can be performed on it. The system type will determine which mathematical constraints on such probabilities make $\Psi$ a valid state and which physical operations we can effect on it.
A complete representation of the state of the system may be achieved by listing the probability of the outcomes for measurements that belong to a so-called `fiducial set', which sometimes has a finite number of elements \cite{lluis11}. For example, the (unnormalized) state of a $d$-dimensional system in quantum theory can be fully described by the vector of probabilities for the $d^2$ outcomes of any tomographically complete measurement. The set of possible states $\mathcal{S}$ of a given system type is always convex. This reflects the fact that any convex combination of two different state preparations is also a valid preparation and should give the corresponding convex combination of measurement statistics.

An \textit{effect} $\mathbf{e}$ is a linear functional on $\mathcal{S}$ that maps each state onto a probability, i.e. a real number between $0$ and $1$. The set of linear functionals with that property form the set $\mathcal{S}^*$, the dual of $\mathcal{S}$. Note that all dichotomic measurements on the system correspond to an element of $\mathcal{S}^*$, when we interpret $\mathbf{e}(\Psi)$ as the probability that the measurement returns outcome $1$. Analogously, any $d$-outcome measurement can be specified by a collection of $d$ effects $\mathbf{e}_j$ such that $\sum_{j=1}^d \mathbf{e}_j(\Psi)=1$ for all valid states $\Psi$. The probability of obtaining outcome $a$ when that measurement is performed on $\Psi$ is given by $\mathbf{e}_a(\Psi)$. In general quantum theory, effects correspond to self-adjoint operators $M$ with $0\leq M\leq \id$.

Note that not all elements of $\mathcal{S}^*$ are required to be allowed effects in the theory. Indeed, the set of physically allowed effects $\mathcal{E}$ may be a strict  {subset} of $\mathcal{S}^*$. 
A theory in which all elements of $\mathcal{S}^*$ \textit{are} allowed effects is called `dual'. The property of duality is often assumed as a starting point in derivations of quantum theory \cite{lluis11, lucien01, lluis13, markus15, oeckl16}, and is usually referred to as \textit{the no-restriction hypothesis} \cite{giulio2}. This hypothesis can be shown to follow from the existence of orthonormal bases for the state space and the equivalence of such bases under reversible transformations \cite{barnum14}. 

\textit{GPT composition rules and normalized Bell inequalities.--} In any complete GPT, given two or more independent systems (not necessarily of the same type), there must exist  {a rule to assign a type to the joint or composite system. In quantum mechanics, for instance, the type of a system composed by a qubit ($\C^2$) and a qutrit ($\C^3$) is a $6$-dit ($\C^2\otimes\C^3=\C^6$)}. In quantum theory, given two systems $A,B$ with state spaces $S_A,  S_B$, embedded in the vector spaces $V_A,V_B$, the state space $S_{AB}$ of the joint system lives in $V_A\otimes V_B$. This is due to the fact that quantum theory satisfies the postulate of \emph{local tomography}, namely, the requirement that local measurements are enough to precisely determine the state of a composite system. In a GPT where local tomography does not hold, the composite state of systems $A$ and $B$ would be given by a vector of the form $r=s\oplus t$, where $s\in V_A\otimes V_B$ is the set of state parameters accessible via local measurements and $t\in V^h_{AB}$ is a set of further parameters only accessible via global or holistic measurements. In such theories, given a number of subsystems with state spaces $V_i$, we define the \emph{locally accessible state} $\Psi_L$ as the projection of the joint state $\Psi$ on the vector space $\bigotimes_i V_i$.

In a given GPT ${\cal T}$, the behavior $p(a_1,...,a_n|x_1,...,x_n)$ generated by $n$ distant parties must be of the form

\be
p(a_1,...,a_n|x_1,...,x_n)=\mathbf{e}_{a_1|x_1}^{1}\otimes...\otimes \mathbf{e}^{n}_{a_n|x_n}(\Psi_L),
\ee

\noindent where $\{\mathbf{e}_{a_k|x_k}^{k}\}\subset \mathcal{E}^*$ are arbitrary valid measurements defined on arbitrary state spaces $V_k$ (not necessarily describing the same system type) in ${\cal T}$ and $\Psi$ is a valid state of ${\cal T}$ for the joint system $A_1$-$A_2$-$...$-$A_n$. The closure of the set of all such behaviors is the \emph{set of correlations of ${\cal T}$}.  The set of correlations of ${\cal T}$ is convex, as parties can produce mixtures of different correlations using shared randomness.

Since the set of correlations of a given theory is convex, we can characterize it via linear inequalities, also known as \emph{generalized Bell inequalities}, or just Bell inequalities, for short. 
Any Bell inequality may be re-scaled into a normalized linear functional $W$ that, when evaluated on any behavior $p(a,b|x,y)$ achievable in the GPT, returns a number between $0$ and $1$, e.~g.:

\be
0\leq W(p):=\sum_{x,y,a,b}W(a,b,x,y)p(a,b|x,y)\leq 1.
\label{norm_Bell}
\ee

\noindent We call such maps \emph{normalized Bell functionals} (NBFs).

Let ${\cal T}$ be a GPT and let $W(a,b,x,y)$ be an NBF in ${\cal T}$. By definition, for any two state spaces, $\cS_A,\cS_B$ in ${\cal T}$, and valid measurements $\{\mathbf{e}_{a|x}\}\subset \cE_A$, $\{\mathbf{f}_{b|y}\}\subset \cE_B$, the vector $\mathbf{W}\equiv\sum_{x,y,a,b}W(a,b,x,y)\mathbf{e}_{a|x}\otimes \mathbf{f}_{b|y}\oplus \vec{0}_{V^h_{AB}}$ belongs to $\cS_{AB}^*$ (note that, when acting on a state, $\mathbf{W}$ taps only the local degrees of freedom). Consequently, if ${\cal T}$ satisfies the no-restriction hypothesis, then $\mathbf{W}$ represents a valid effect on the joint system $A$-$B$, i.e.~$\mathbf{W} \in \cE_{AB}$. Any NBF can thus be understood as a map that turns a collection of local measurements into a new joint effect $\mathbf{W}$, which we call the Bell effect. In any GPT, this effect can always be completed into a two-outcome measurement simply by combining it with another effect that sums up to a normalisation factor, defining what we call, somehow analogously to the quantum case a \emph{Bell measurement}.  {Bell measurements may have more than two effects}, i.e., they may be constructed from a collection of NBFs $\{W_\alpha\}_\alpha$ with the property that $\sum_\alpha W_\alpha (p)=1$ for all achievable behaviors $p$. Intuitively, $\alpha$ denotes the `outcome' of the NBF. We say that a set of NBFs satisfying this property is complete. In the rest of the paper, it will be enough to consider two-outcome Bell measurements to prove our claim. 

Remarkably, and this is crucial for what follows, in order to compute the probability of obtaining the outcome $\alpha=\beta$ when we apply a Bell measurement (constructed from a complete set of NBFs $\{W_\alpha\}_\alpha$ over measurements $\{\mathbf{e}_{a|x}\}$, $\{\mathbf{f}_{b|y}\}$) on a bipartite system  we do not need to know anything about the underlying GPT theory. In other words we do not need to know about $\{\mathbf{e}_{a|x}\}$, $\{\mathbf{f}_{b|y}\}$ or $\Psi$: it suffices to know the probabilities $p(a,b|x,y)=\mathbf{e}_{a|x}\otimes \mathbf{f}_{b|y}(\Psi_L)$.

\begin{figure}
\begin{center}
\subfigure[\, The family $U$ of NBFs $\{U_{\alpha|\xi}\}$ that defines a collection of Bell measurements.]{
\begin{tikzpicture}[scale=0.5]
\shade[draw, thick, ,rounded corners, inner color=white,outer color=gray!50!white] (-2,-1) rectangle (0,1) ;
\draw[thick] (-1,-1) -- (-1,-1.5);
\draw[thick] (-1,1.5) -- (-1,1);
\node at (-1,-1.7) {$a$};
\node at (-1,1.7) {$x$};

\shade[draw, thick, ,rounded corners, inner color=white,outer color=gray!50!white, xshift=2cm] (1,-1) rectangle (3,1) ;
\draw[thick, xshift=2cm] (2,-1) -- (2,-1.5);
\draw[thick, xshift=2cm] (2,1.5) -- (2,1);
\node[xshift=1cm] at (2,-1.7) {$b$};
\node[xshift=1cm] at (2,1.7) {$y$};

\draw[dotted, thick, rounded corners] (-2.2,-1.2) rectangle (5.2, 1.2);
\node at (1.5,0) {$U_{\alpha|\xi}$};
\node at (6,0) {$\equiv$};

\shade[draw, thick, ,rounded corners, inner color=white,outer color=gray!50!white] (7,-1) rectangle (9,1) ;
\draw[thick] (8,-1) -- (8,-1.5);
\draw[thick] (8,1.5) -- (8,1);
\node at (8,-1.7) {$\alpha$};
\node at (8,1.7) {$\xi$};
\node at (8,0) {$U$};
\end{tikzpicture}}
\vskip 0.3cm
\subfigure[\, An NBF $V$ in a bipartite scenario.]{
\begin{tikzpicture}[scale=0.5]
\shade[draw, thick, ,rounded corners, inner color=white,outer color=gray!50!white] (-2,-1) rectangle (0,1) ;
\draw[thick] (-1,-1) -- (-1,-1.5);
\draw[thick] (-1,1.5) -- (-1,1);
\node at (-1,-1.7) {$\alpha$};
\node at (-1,1.7) {$\xi$};

\shade[draw, thick, ,rounded corners, inner color=white,outer color=gray!50!white, xshift=2cm] (1,-1) rectangle (3,1) ;
\draw[thick, xshift=2cm] (2,-1) -- (2,-1.5);
\draw[thick, xshift=2cm] (2,1.5) -- (2,1);
\node[xshift=1cm] at (2,-1.7) {$c$};
\node[xshift=1cm] at (2,1.7) {$z$};

\draw[dotted, thick, rounded corners] (-2.2,-1.2) rectangle (5.2, 1.2);
\node at (1.5,0) {$V$};
\node at (6,0){};
\node at (-3.5,0) {};
\end{tikzpicture}}
\vskip 0.3cm
\subfigure[\, Tripartite NBF $W$ that effectively arises when identifying the left device in the bipartite scenario (b) with the family of Bell measurements
$U$ defined by the bipartite NBFs $\{U_{\alpha|\xi}\}$.]{
\begin{tikzpicture}[scale=0.5]
\shade[draw, thick, ,rounded corners, inner color=white,outer color=gray!50!white] (-2,-1) rectangle (0,1) ;
\draw[thick] (-1,-1) -- (-1,-1.5);
\draw[thick] (-1,1.5) -- (-1,1);
\node at (-1,-1.7) {$a$};
\node at (-1,1.7) {$x$};

\shade[draw, thick, ,rounded corners, inner color=white,outer color=gray!50!white, xshift=2cm] (1,-1) rectangle (3,1) ;
\draw[thick, xshift=2cm] (2,-1) -- (2,-1.5);
\draw[thick, xshift=2cm] (2,1.5) -- (2,1);
\node[xshift=1cm] at (2,-1.7) {$b$};
\node[xshift=1cm] at (2,1.7) {$y$};

\shade[draw, thick, ,rounded corners, inner color=white,outer color=gray!50!white, xshift=4cm] (4,-1) rectangle (6,1) ;
\draw[thick, xshift=4cm] (5,-1) -- (5,-1.5);
\draw[thick, xshift=4cm] (5,1.5) -- (5,1);
\node[xshift=2cm] at (5,-1.7) {$c$};
\node[xshift=2cm] at (5,1.7) {$z$};

\draw[dotted, thick, rounded corners] (-2.2,-1.2) rectangle (10.2, 1.2);
\node at (-4,0) {$W$};
\node at (-3,0) {$\equiv$};
\node at (11,0){};

\draw [thick, decorate,decoration={brace,amplitude=10pt,aspect=0.47}] (-2.2,2)--(5.5,2);
\shade[draw, thick, ,rounded corners, inner color=white,outer color=gray!50!white] (0.5,4) rectangle (2.5,6) ;
\draw[thick] (1.5,4) -- (1.5,3.5);
\draw[thick] (1.5,6.5) -- (1.5,6);
\node at (1.5,3.3) {$\alpha$};
\node at (1.5,6.7) {$\xi$};
\node at (1.5,5) {$U$};

\end{tikzpicture}}
\end{center}
\caption{Composition of NBFs. (a) A family of bipartite NBFs, such that $\{U_{\alpha|\xi}\}_\alpha$ defines a Bell measurement for each $\xi$. This collection of Bell measurements defines from the original bipartite box $p(ab|xy)$ a new device with input $\xi$ and output $\alpha$, where $p(\alpha|\xi) = \sum_{a,b,x,y} U_{\alpha|\xi}(a,b,x,y) \, p(a,b,x,y)$. (b) A bipartite NBF $V$ acting on a box $p(\alpha,c | \xi,z)$. (c) An implementation of the device in (b) where now the left-hand-side box is realised by the device $U$ of (a). Testing the NBF $V$ in this realisation of the box $p(\alpha,c | \xi,z)$ effectively tests a tripartite inequality given by the coefficients $W(a,b,c,x,y,z)=\sum_{\alpha,\xi}V(\alpha,c,\xi,z)U_{\alpha|\xi}(a,b,x,y)$. Should the GPT satisfy the no-restriction hypothesis and $\{U_{\alpha|\xi}\}_\alpha$ and $V$ be AQ NBFs, $W$  {would be an AQ NBF}.}
  \label{wired}
\end{figure}
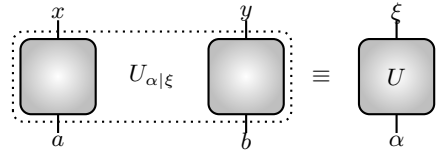
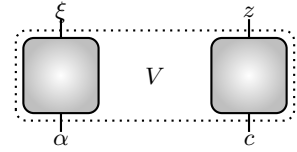
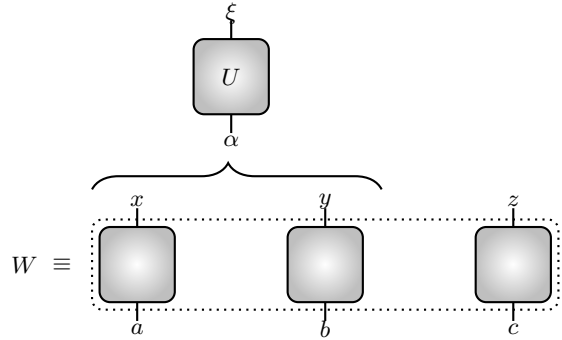

Now, since under the no-restriction hypothesis  a set of GPT measurements  $\{\mathbf{e}_{a|x}\}$, $\{\mathbf{f}_{b|y}\}$ can be combined using an NBF to define a new valid effect itself, then we can act on it with another NBF. This allows us to iterate the process and generate new Bell measurements from others just by contracting setting and outcome indices, see Figure \ref{wired}. 

For instance, let $\{\{U_{\alpha|\xi}\}_\alpha\}_\xi$ be a collection of bipartite NBFs labeled by the ``setting" $\xi$ and the ``outcome'' $\alpha$. For each  $\xi$, $\{U_{\alpha|\xi}\}_\alpha$ is a complete set of NBFs. Then, under the no-restriction hypothesis, for any state spaces $\cS_A,\cS_B$ and measurements $\{\mathbf{e}_{a|x}\}\subset \cS_A^*,\{\mathbf{f}_{b|y}\}\subset \cS_B^*$, for every $\xi$, the vectors $\mathbf{u}_{\alpha|\xi}\equiv \sum_{a,b,x,y}U_{\alpha|\xi}(a,b,x,y)\mathbf{e}_{a|x}\otimes\mathbf{f}_{b|y}\oplus 0_{V_{AB}}$ represent measurements with outcomes labeled by $\alpha$. Now let $V$ be another bipartite NBF. Again invoking the no-restriction hypothesis, we have that for any state space $\cS_C$ and measurements $\{\mathbf{g}_{c|z}\}\subset \cS_C^*$, $\mathbf{V}(\{\mathbf{u}_{\alpha|\xi}\},\{\mathbf{g}_{c|z}\})$ must be a valid effect. In terms of $\mathbf{e}_{a|x},\mathbf{f}_{b|y}$, this effect corresponds to the vector $\sum_{a,b,c,x,y,z}W(a,b,c,x,y,z)\mathbf{e}_{a|x}\otimes\mathbf{f}_{b|y}\otimes\mathbf{g}_{c|z}\oplus \mathbf{0}_{V^h_{rest}}$, where $V^h_{rest}$ denotes the space of parameters of the composite system $A$-$B$-$C$ not accessible via local tomography, and

\be
W(a,b,c,x,y,z)\equiv\sum_{\alpha,\xi}V(\alpha,c,\xi,z)U_{\alpha|\xi}(a,b,x,y).
\label{composition}
\ee
\noindent  {But now, since $\cS_A,\cS_B,\cS_C$ and $\mathbf{e}_{a|x}, \mathbf{f}_{b|y}, \mathbf{g}_{c|z}$ are arbitrary, it follows that $W(a,b,c,x,y,z)$ defines a tripartite NBF. Thus, we see that the no-restriction hypothesis implies a rich structure in the set of possible joint measurements in the theory given only its correlation structure. Also, it implies a set of consistency constraints among the theory's NBFs.}

\textit{The almost quantum set is incompatible with the no-restriction hypothesis.--} To prove our main result, we combine different AQ bipartite NBFs as in eq.~\eqref{composition}, into a new functional that should be a valid NBF for AQ tripartite correlations should the No Restriction Hypothesis hold.  The contradiction comes when the resulting inequality is violated by AQ correlations. 
To find the NBFs, we used a see-saw-type method \cite{see_saw} that rested upon the SDP characterizations of both the set of AQ correlations ~\cite{almost} and that of AQ NBFs, see the Supplemental Material.
 
Our first NBFs for AQ correlations are defined in the $(2,3,2)$ Bell scenario and read
\begin{align}
U_{0|0}(p)= & 1-p_A(0|1)-p_B(0|1)+2p(00|11),\nonumber\\
U_{0|1}(p)= & 1-0.0329p_B(0|0)-0.7117p_B(0|2)\nonumber\\
&-0.0329p_A(0|0) - 0.8418p(0,0|0,0) \nonumber\\
&+0.6359 p(0,0|0,2)-0.7117 p_A(0|2)\nonumber\\
&+0.6359 p(0,0|2,0)+0.4360 p(0,0|2,2).
\label{Us_NBF}
\end{align}
\noindent $U_{0|0}$ is actually a wiring \cite{closed}: namely, it corresponds to measuring setting $1$ in both systems and outputting $0$ when the outcomes are equal. 
$U_{0|1}$, on the contrary, can be shown not to be wirings or a convex combination thereof. However, one can verify by running the SDP provided in~\cite{almost} that it is an AQ  {NBF}.

Similarly, one can check that the $(2,2,2)$ Bell functional

\begin{align}
V(p)=&0.1590+0.8372 p_B(0|0) +0.0031p_B(0|1)\nonumber\\
&-0.1544 p_A(0|0)-0.6132 p(0,0|0,0)\nonumber\\
& +0.5547p(0,0|0,1) +0.5884p_A(0|1)\nonumber\\
 &-0.5902 p(0,0|1,0)-0.7404 p(0,0|11).
\label{V_NBF}
\end{align}
\noindent is an AQ NBF.

Finally, consider the composition $W$ of these NBFs according to (\ref{composition}). Minimizing the value $W(p)$ over all AQ tripartite correlations $p$, we find $\min_{p\in\tilde{Q}}W(p)\approx -0.0033$. Since, by definition, all normalized Bell inequalities satisfy $W(p)\geq 0$ for all $p\in\tilde{Q}$, we conclude that $W$ is not a normalized Bell inequality. Hence, the AQ set cannot reflect the correlations of any GPT respecting the no-restriction hypothesis.

{Note that this result is robust, i.e., it can be extended to prove that the set of tripartite correlations of any GPT satisfying the no-restriction hypothesis cannot even approximate the almost quantum set. 
Indeed, let $p$ be the almost quantum distribution maximally violating $W$, and let $\tilde{Q}'$ be a set of correlations in the $(3,3,2)$ Bell scenario that approximates the almost quantum one. 
Then, there exist a distribution $p'\in\tilde{Q}'$ and NBFs $V'$ and $\{U'_{0|x}\,, x=0,1\}$ for $\tilde{Q}'$ such that $p'\approx p$, $V'\approx V, U'_{0|x}\approx U_{0|x}, x=0,1$. Composing $V'$ and $U'_{0|x}$ according to eq.~\eqref{composition}, we obtain the Bell functional $W'$. Should the GPT behind $\tilde{Q}'$ satisfy the No Restriction Hypothesis, $W'$ must be an NBF for $\tilde{Q}'$. However, if $\tilde{Q}$ and $\tilde{Q}'$ are close enough, $W'(p')\approx -0.0033$, contradicting the claim.}

\textit{Conclusions.--} In this paper we have addressed the question of whether a theory that predicts AQ correlations can satisfy the no-restriction hypothesis. We have answered this question in the negative. To do so, we have shown that the no-restriction hypothesis implies a series of non-trivial compatibility constraints among the NBFs satisfied by a set of correlations. We then went on to show that NBFs satisfied by AQ correlations violate these compatibility constraints. From a more general perspective, our formalism provides a way to test the validity of the no-restriction hypothesis in a device-independent way.

Our work opens two main research questions that deserve further investigation. First, it would be interesting to identify GPT theories predicting AQ correlations and the {constraints they impose on the set of measurements}. 
Second, one may wonder which sets of correlations are compatible with the no-restriction hypothesis. General non-signalling and quantum correlations are examples of those sets. How to identify other such sets is left for future work. In particular, leaving aside general non-signalling correlations, is there any set of correlations that is strictly larger than the quantum set and does not violate the no-restriction hypothesis?

\texttt{Acknowledgements.--} We thank Tomer Barnea, Lucien Hardy, Markus M\"uller, John Selby, Tony Short and Rob Spekkens for fruitful discussions. 
We acknowledge support from the European Research Council (CoG QITBOX), an Axa Chair in Quantum Information Science, Spanish MINECO (QIBEQI FIS2016-80773-P and Severo Ochoa SEV-2015-0522), Fundaci\'{o} Privada Cellex, and Generalitat de Catalunya (CERCA Program  and SGR1381).
A.A. and M.N. acknowledge support from FQXi via the program ``The physics of what happens''.
YG  acknowledges  funding  from  the  Austrian  Science Fund (FWF) through the START project Y879-N27.
This research was supported in part by Perimeter Institute for Theoretical Physics. Research at Perimeter Institute is supported by the Government of Canada through the Department of Innovation, Science and Economic Development Canada and by the Province of Ontario through the Ministry of Research, Innovation and Science.

\appendix

\section*{Supplemental Material : Almost quantum normalised Bell Functionals}

 {In this appendix we introduce a semidefinite programming characterisation of almost quantum NBFs and explain how we used it to identify the counterexample to the composition rule (4) given by eqs. (5) and (6).}

Due to the close connection between quantum and almost quantum correlations, it will be convenient to use linear combinations of ``symbolic'' projectors, $\{E_{a|x}\}_{a,x}$ for Alice and $\{F_{a|y}\}_{b,y}$ for Bob (and products thereof), in order to express almost quantum NBF. These symbolic projectors satisfy the same algebraic relations as the operators defining quantum measurements over a bipartite system, namely:

\begin{align}
&E_{a|x}=(E_{a|x})^2=(E_{a|x})^\dagger, F_{b|y}=(F_{b|y})^2=(F_{b|y})^\dagger,\nonumber\\
&\sum_aE_{a|x}=\sum_bF_{b|y}=1, [E_{a|x},F_{b|y}]=0.
\label{identities}
\end{align}

\noindent For instance, the product $E_{a|x}F_{b|y}$ denotes the NBF $0\leq p(a,b|x,y)\leq 1$. A general bipartite NBF would hence be expressed as $\sum_{x,y,a,b}W(a,b,x,y)E_{a|x}F_{b|y}$.

NBFs in the almost quantum set possess a very simple structure: they correspond to Sums of Hermitian Squares (SOS) of certain kind of polynomials of those symbolic quantum measurement operators. Namely, $W(a,b,x,y)$ is an almost quantum NBF iff it can be expressed as a \emph{sum of Hermitian squares}, i.e., if there exist Bell functionals $\{f_i\}_i$, $\{g_i\}_i$ in operator form such that

\begin{align}
&\sum_{x,a,y,b}W(a,b,x,y)E_{a|x}F_{b|y}=\sum_i f_i^\dagger f_i,\nonumber\\
&1-\sum_{x,a,y,b}W(a,b,x,y)E_{a|x}F_{b|y}=\sum_i g_i^\dagger g_i.
\label{SOS}
\end{align}
The existence of such decompositions can be established via semidefinite programming \cite{yeong-toner}. To prove that, we first consider the related problem of minimizing the value of a Bell functional $W$ over the set of almost quantum correlations. 

Let the outcomes $a,b$ run from $1$ to $d$. Due to relations (\ref{identities}), a basis of independent operators to express NBFs is given by the monomials $M=\{1\}\cup \{E_{a|x}, F_{b|y}:a,b\not=d\}\cup \{E_{a|x}F_{b|y}:a,b\not=d\}$, which in the following we denote by $\{\Pi_{\gamma}\}_{\gamma}$. From now on we will express $W$ in basis $M$, i.e., $\{W(a,b,x,y):a,b,x,y\}\to \{W(\gamma):\Pi_{\gamma}\in M\}$. Following \cite{almost}, we have that the minimization of linear functionals $W(\gamma)$ over the almost quantum set can be cast as a semidefinite program:

\begin{align}
&\min\sum_{\gamma\in M}W(\gamma)\Gamma(\gamma,1),\nonumber\\
\mbox{s.t. }&\Gamma\geq 0,\Gamma(1,1)=1;\nonumber\\
&\Gamma(\gamma,\gamma')=\Gamma(\gamma'',\gamma')\delta_{c,c'},\nonumber\\
&\mbox{for } \Pi_{\gamma}=E_{c|z}s,
	\quad \Pi_{\gamma'}=E_{c'|z}t,
		\quad \Pi_{\gamma''}=s,\nonumber\\ 
&\mbox{or }\;\Pi_{\gamma}=sF_{c|z}, 
	\quad \Pi_{\gamma'}=tF_{c'|z},
		\quad \Pi_{\gamma''}=s;\nonumber\\
&\Gamma(\gamma,\gamma')=\Gamma(\gamma,\gamma'')\delta_{c,c'},\nonumber\\
&\mbox{for }\Pi_{\gamma}=E_{c|z}s, 
	\quad \Pi_{\gamma'}=E_{c'|z}t, 
		\quad \Pi_{\gamma''}=t,\nonumber\\
&\mbox{or }\;\Pi_{\gamma}=sF_{c|z},
	\quad \Pi_{\gamma'}=tF_{c'|z}, 
		\quad \Pi_{\gamma''}=t\; .
\label{primal}
\end{align}
\noindent Here $\Gamma$ is a matrix whose columns (in a certain basis) are labeled by elements of the basis $M$ of symbolic monomials. 
That is, $\Gamma=\sum_{\gamma,\gamma'}\Gamma(\Pi_{\gamma},\Pi_{\gamma'})\ket{\gamma}\bra{\gamma'}$, where $\{\ket{\gamma}\}_{\gamma}$ is an orthonormal basis. $E_{a|x}s,E_{a'|x}t$, ($sF_{b|y},tF_{b'|y}$) denote monomials in $M$, and hence $s,t$ must equal to either $1$ or $F_{b|y}$ for some $b,y$ ($E_{a|x}$ for some $a,x$). The SDP dual of this problem is \cite{sdp}:

\begin{align}
&\max\lambda,\nonumber\\
\mbox{s.t. } & \nonumber \\ &Z\equiv\sum_{\gamma\in M}W(\gamma)\ket{\gamma}\bra{1}-\lambda\ket{1}\bra{1}+\sum_{i}c_iG_i\geq 0. \label{e:Z}
\end{align}
Here the matrices $\{G_i\}$ exhaust all the matrices of the form:
\be
\ket{\gamma}\bra{\gamma'}-\ket{\gamma''}\bra{\gamma'}\delta_{c,c'},
\ee
\noindent for $\Pi_{\gamma}=E_{c|z}s$, $\Pi_{\gamma'}=E_{c'|z}t$, $\Pi_{\gamma''}=s$ (and $\Pi_{\gamma}=sF_{c|z}$, $\Pi_{\gamma'}=tF_{c'|z}$, $\Pi_{\gamma''}=s$),
and

\be
\ket{\gamma}\bra{\gamma'}-\ket{\gamma}\bra{\gamma''}\delta_{c,c'},
\ee
\noindent for $\Pi_{\gamma}=E_{c|z}s$, $\Pi_{\gamma'}=E_{c'|z}t$, $\Pi_{\gamma''}=t$ (and $\Pi_{\gamma}=sF_{c|z}$, $\Pi_{\gamma'}=tF_{c'|z}$, $\Pi_{\gamma''}=t$).

If we multiply the matrix $Z$ in eq.~\eqref{e:Z} on both sides by the vector of symbolic monomials $\ket{u}\equiv\sum_{\gamma}\Pi_{\gamma}\ket{\gamma}$, we obtain an element of $\mbox{lin}(M)$ equivalent to the polynomial
\begin{equation}
-\lambda +\sum_{\gamma\in M}W(\gamma)\Pi_{\gamma}
\end{equation}
The matrices $\{G_i\}$ can be interpreted as \emph{substitution rules} of the form $E_{a|x}sE_{a'|x}t\to E_{a|x}st\delta_{aa'}$ (for $E_{a|x}s,E_{a'|x}t\in M$) to transform the matrix representation of the polynomial above. Since the final matrix $Z$ is positive semidefinite, it can be expanded as $Z=\sum_i \ket{f_i}\bra{f_i}$, and hence
\begin{align}
-\lambda +\sum_{\gamma\in M}W(\gamma)\Pi_{\gamma}=\sum_i \braket{u}{f_i}\braket{f_i}{u}
\end{align}
and so $-\lambda +\sum_{\gamma\in M}W(\gamma)\Pi_{\gamma}$ is a Sum Of Hermitian Squares (SOS) of linear combinations $\braket{u}{f_i}$ of elements in $M$.

Under the assumption that there is no duality gap between primal and dual problems, this means that $\sum_{a,b,x,y} W(a,b,x,y)p(a,b|x,y)\geq 0$ for all almost quantum distributions iff $\sum_{a,b,x,y}W(a,b,x,y)E_{a|x}F_{b|y}$ admits such an SOS decomposition. If we further demand that $\sum_{a,b,x,y} W(a,b,x,y)p(a,b|x,y)\leq 1$, then the polynomial $1-\sum_{a,b,x,y}W(a,b,x,y)E_{a|x}F_{b|y}$ must also admit a similar SOS decomposition. We hence arrive at eq. (\ref{SOS}).

It remains to be seen that the semidefinite program (\ref{primal}) does not exhibit a duality gap. This can be seen by assigning a Hilbert space $\cH_x$ ($\cH_y$) of dimension $d$ to each one of  Alice's (Bob's) measurement settings, such that the total space is $\mathcal{H}_A\otimes \mathcal{H}_B = \bigotimes_{x=1}^d\mathcal{H}_x \otimes \bigotimes_{y=1}^d\mathcal{H}_y $. Next  take $E_{a|x}=\proj{a}_x\otimes \id_{\bar{x}}\otimes \id_B$ ($F_{b|y}=\id_A\otimes\proj{b}_y\otimes \id_{\bar{y}}$), where $\bar{x}$ ($\bar{y}$) denotes the tensor product of all Hilbert spaces apart from $\cH_x$ ($\cH_y$) and $A$ ($B$) denotes the tensor product of all of Alice's (Bob's) Hilbert spaces. By defining $\Gamma(\gamma,\gamma')=\frac{1}{d_Ad_B}\mbox{tr}(\Pi_{\gamma}\Pi_{\gamma'})$, it is easy to demonstrate that the matrix $\Gamma$ is positive \emph{definite} and satisfies all the linear constraints in (\ref{primal}). Hence the primal problem has strictly feasible points and therefore there is no duality gap between the two problems \cite{sdp}.

 {
In order to identify the NBFs (5) and (6), we considered minimizing the quantity

\be
\sum_{a,b,c,x,y,z}W(a,b,c,x,y,z)p(a,b,c|x,y,z),
\label{figure_of_merit}
\ee
\noindent over all almost quantum distributions $p(a,b,c|x,y,z)$ and Bell functionals $W(a,b,c,x,y,z)$ of the form (4), with $U_{\alpha|\xi},V$ being almost quantum NBFs. Note that the figure of merit (\ref{figure_of_merit}) is multilinear in the variables of the problem $p,U,V$. Hence, for fixed values of two of those variables, one can optimize the third one via semidefinite programming. This is exactly what we did: starting with a guess on $V$ and random values for $U$, we optimized over almost quantum distributions $p$. Then we optimized over $U$, keeping $p$ and $V$ fixed, and then over $V$, with $p$ and $U$ fixed. Iterating this see-saw scheme \cite{see_saw}, that we had to re-initiate a few times with different random seeds for $U$, we managed to reach a negative value for (\ref{figure_of_merit}), hence proving that the composition rule (4) does not hold for almost quantum NBFs.}


\begin{thebibliography}{99}
\bibitem{popescu}
S. Popescu and D. Rohrlich, Foundations of Physics {\bf24} (3), 379-385 (1994).
\bibitem{tsirel_box}
L.A. Khalfin, B.S. Tsirelson, \emph{Quantum and quasi-classical analogs of Bell inequalities}, Symposium on the Foundations of Modern Physics 1985 (ed. Lahti et al.; World Sci. Publ.), 441-460 (1985).
\bibitem{jeanclaude}
W. van Dam, arXiv:quant-ph/0501159.
\bibitem{brassard}
G. Brassard, H. Buhrman, N. Linden, A. A. Methot, A. Tapp and F. Unger, F., Phys. Rev. Lett., \textbf{96} 250401, (2006).
\bibitem{linden}
N. Linden, S. Popescu, A. J. Short, and A. Winter, Phys. Rev. Lett. \textbf{99}, 180502 (2007).
\bibitem{marcin}
M. Pawlowski, T. Paterek, D. Kaszlikowski, V. Scarani, A. Winter, and M. Zukowski, Nature \textbf{461}, 1101 (2009).
\bibitem{mac_loc}
M. Navascu\'es and H. Wunderlich, Proc. Royal Soc. A \textbf{466}, 881-890 (2009).
\bibitem{loc_orth}
T. Fritz, A. B. Sainz, R. Augusiak, J. B. Brask, R. Chaves, A. Leverrier and A. Ac\'in, Nature Comm \textbf{4}, 2263 (2013).
\bibitem{almost}
M. Navascu\'es, Y. Guryanova, M. J. Hoban and A. Ac\'in, Nat. Comm. \textbf{6}, 6288 (2015).

\bibitem{sdp}
Vandenberghe, L. and Boyd, S. SIAM Review \textbf{38}, 49 (1996).

\bibitem{lluis11} Ll. Masanes and M. P. Mueller, New Journal of Physics \textbf{13}, 063001 (2011).
\bibitem{lucien01} L. Hardy, (2001) preprint at arXiv:quant-ph/0101012.
\bibitem{lluis13} Ll. Masanes, M. P. Mueller, R. Augusiak, D. Perez-Garcia, PNAS \textbf{110}(41), 16373 (2013).
\bibitem{markus15} M. P. Mueller and Lluis Masanes, book chapter in"Quantum Theory: Informational Foundations and Foils", G. Chiribella and R. Spekkens eds., Springer (2016). Preprint at arXiv:1203.4516.
\bibitem{oeckl16} R. Oeckl, (2016) preprint at arXiv:1610.09052.
\bibitem{giulio2} G. Chiribella, G. M. D'Ariano and P. Perinotti, Phys. Rev. A {\bf 81}, 062348 (2010).
\bibitem{barnum14} H. Barnum, M. P. Mueller and C. Ududec, New Journal of Physics \textbf{16}, 123029 (2014).
\bibitem{yeong-toner}
A. C. Doherty, Y.-C. Liang, B. Toner and S. Wehner, Proceedings of IEEE Conference on Computational Complexity, 199--210 (2008). 
\bibitem{closed}
J. Allcock, N. Brunner, N. Linden, S. Popescu, P. Skrzypczyk and T. V\'ertesi, Phys. Rev. A \textbf{80}, 062107 (2009).
\bibitem{see_saw}
K.F. P\'al, T. V\'ertesi, Phys. Rev. A {\bf 82}, 022116 (2010).
\end{thebibliography}
\end{document}